\pdfoutput=1
\documentclass{article}

\usepackage[nonatbib,preprint]{nips_2018}

\usepackage[utf8]{inputenc} 
\usepackage[T1]{fontenc}    
\usepackage{hyperref}       
\usepackage{url}            
\usepackage{booktabs}       
\usepackage{amsfonts}       
\usepackage{nicefrac}       
\usepackage{microtype}      
\usepackage{graphicx}
\usepackage{blindtext}
\usepackage{color}

\title{3D Deep Learning with voxelized atomic configurations for modeling atomistic potentials in complex solid-solution alloys}

\author{
Rahul Singh*\\
Iowa State University
\And
Aayush Sharma*\\
Iowa State University
\And
Onur Rauf Bingol*\\
Iowa State University
\And
Aditya Balu\\
Iowa State University
\AND
Ganesh Balasubramanian\\
Lehigh University
\And
Duane D. Johnson\\
Iowa State University
\And
Soumik Sarkar\\
Iowa State University
\\
\thanks{Equal Contribution}}

\begin{document}

\maketitle

\begin{abstract}
  The need for advanced materials has led to the development of complex, multi-component alloys or solid-solution alloys. These materials have shown exceptional properties like strength, toughness, ductility, electrical and electronic properties. Current development of such material systems are hindered by expensive experiments and computationally demanding first-principles simulations. Atomistic simulations can provide reasonable insights on properties in such material systems. However, the issue of designing robust potentials still exists. In this paper, we explore a deep convolutional neural-network based approach to develop the atomistic potential for such complex alloys to investigate materials for insights into controlling properties. In the present work, we propose a voxel representation of the atomic configuration of a cell and design a 3D convolutional neural network to learn the interaction of the atoms. Our results highlight the performance of the 3D convolutional neural network and its efficacy in machine-learning the atomistic potential. We also explore the role of voxel resolution and provide insights into the two bounding box methodologies implemented for voxelization. 
\end{abstract}

\section{Introduction}

Over the past decade a subset of complex solid-solution alloys (CSAs), known as high-entropy alloys (HEAs), have captured the imagination of researchers across the globe. HEAs have demonstrated remarkable mechanical properties including, but not limited to, enhanced structural strength, as well as improved resistance to fatigue, oxidation, corrosion, and wear~\cite{Yeh2004,Singh2018,Gao2018,Gorsse2017,MiracleSenkov2017}. These findings have encouraged extensive research to explore the potential of HEAs for high temperature applications and for engineering systems operating under harsh environments~\cite{Yeh2004,MiracleSenkov2017,Miracle2015,Miracle2017}, and hence, the need to understand the material nano- and micro-structures. The infinite permutations and combinations of the atomic species in varying concentrations that can be employed to produce CSAs, creates an enormous challenge in designing alloys with novel compositions and structures. The plausible inclusion of several different elements in notable proportions to synthesize HEAs enables a high dimensional design space explorations relative to traditional alloys~\cite{Singh2018}. The design space is impractically large for exploration solely by experimental methods or high-performance atomistic simulations. A majority of the investigations have resorted to resource intensive experimental measurements or first-principles calculations, while computational modeling by atomistic simulations, especially for the sub-micron length processes, have suffered due to the absence of well-defined force fields to describe interatomic interactions~\cite{Yeh2004,Sharma2016,Sharma012017,Sharma022017}. Computationally demanding electronic-structure calculations using density functional theory (DFT) are often restricted to simulating a few hundred atoms, for time-scales less than a nanosecond~\cite{Sharma2016,Sharma012017,Sharma022017}, which molecular dynamics (MD) simulations can overcome if suitable potential functions are available to describe the atomic interactions. MD can predict deterministically nano/microscopic phenomena (deformation and motion of dislocations, phase-transition) and predict material properties at a fraction of the computational time compared to first-principles-based techniques~\cite{Sharma2016,Sharma012017,Sharma022017}.\\
An interatomic potential function or potential energy surface (PES) refers to the mathematical equation that provides a direct functional relation between the configurations, positions and potential energy of a group of atoms. Analytical potentials provide a simpler, direct relation between the molecular configurations and its potential energy in closed form, and facilitate a quick energy calculation, but they are usually derived by introducing physical approximations~\cite{Artrith2016}. These potentials represent a necessary compromise between efficiency and accuracy, given that the essential characteristics of the atomic interactions are reasonably described. In MD simulations~\cite{Plimpton1995}, force-field functions, such as the embedded atom method (EAM)~\cite{Zhou2004}, are available for a limited set of atom-combinations and can be employed to describe the self and cross interactions of the different participating species. They ideally should reproduce configurational energies and defect energies that control relevant physical mechanisms, which, at present, is a short-coming of most empirical potentials. 

The vastness of the CSAs composition landscape, and the absence of PES information for several of the elemental combinations, compels the development of a method for deriving useful potential functions for these materials a timely scientific problem to address. Here, we describe a deep-learning (DL) method seeded with data from ab-initio molecular dynamics (AIMD) calculations as a route to developing interatomic potentials. Such potentials essentially utilize ML techniques to construct direct functional correlation between the atomic position, configuration and energy~\cite{Artrith2016}, and employ a consistent set of atomistic data~\cite{Artrith2016,Liu2017,Behler2016}. A robust ML potential enables (a) faster evaluations especially for large MD or Monte Carlo (MC) simulations, (b) general applicability to all possible interactions within a system containing various atomic species, and (c) high predictive accuracy that are comparable to DFT predictions.

\paragraph{Related Work:}
Application of machine learning (ML) to atomistic simulations to predict PES has been explored by several in the past~\cite{Artrith2016, Liu2017, Behler2016}. However, most of these approaches have been performed for pure elemental simulations. Also, because the atomic feature descriptor are volumetric in nature (i.e., the position of atoms, interatomic distances, etc.) and evaluating it in 3D domain of the cell shall makes the ML method more effective. The key novelty in this paper is to use a volumetric representation of the physical descriptors using a voxel grid and thus predicting atomistic potential for multi-component alloys. To best of our knowledge, there are no or very sparse attempts for using voxel-based (3D) convolutional Neural Networks for predicting PES for complex solid solution alloys. 

\paragraph{Contributions:}
Our specific contributions in this work are:
\begin{enumerate}
    \item We produce a potential to describe efficiently and accurately the PES for molecular simulations of equiatomic MoTaW HEA, as a  representative CSA.
    \item We represent the inputs for ML-PES using a voxel representation using the atomic scale information, such as interatomic distances, from computational chemistry calculations that play a central role in the construction of PES~\cite{Artrith2016,Bartok2013}.
    \item We build a framework (called as Machine-Learning-based-Potential Energy Surface, or ML-PES) for using deep convolutional neural networks to learn from ab-initio molecular dynamics (AIMD) simulations of multi-component alloys.

\end{enumerate}
 The flow of the paper is as follows: In section 2, we discuss about Data preparation and voxelization of Ab-Initio Molecular Dynamics data of CSAs. In section~\ref{sec:DNN}, we provide details about the deep convolutional neural network used for machine learning and the training process. In section~\ref{sec:results}, we present the results and discussion.

\begin{figure}[t]
\centering
\includegraphics[width = 0.8\textwidth]{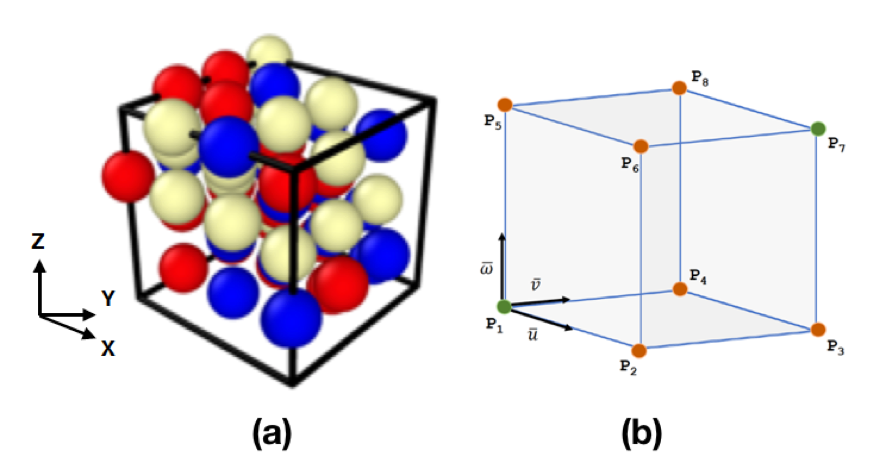}
\caption{(a) Atomic representation of a 54-atom ternary, body centered cubic (BCC) crystal structure that forms the unit cell of an equiatomic MoTaW HEA used as a testbed for developing the ML potential. Atoms have been color coded (Mo: Red, Ta: Blue, W: White) to visualize the local environmental complexity in the alloys. (b) Schematic representation of a volume element (voxel) used for data preparation.}
\label{fig:struct}
\end{figure}


\section{Ab-Initio Molecular Dynamics for CSAs}
\subsection{First-principles-based calculations}
We perform ab-initio molecular dynamics (AIMD) simulations with the VASP 5.4 package\cite{Kresse1993,Kresse1994}, using the Perdew-Burke-Ernzerhof (PBE)\cite{Perdew1996} exchange-correlation functional and the projected-augmented wave (PAW) potentials\cite{Blochl1994,Kresse1999}. The three elements Ta, Mo and W form the equiatomic ternary HEA. The initial structures are assimilated from our in-house hybrid Cuckoo-Search with local environments optimized by Monte-Carlo algorithm that minimizes the short-range order (SRO)\cite{Singh2018}, and represent near random configurations that are desirable for disordered solid solutions (CSAs or HEAs). We employ an energy cut-off of 400 eV and Born-Oppenheimer convergence of 10$^{-6}$ eV at each time step. A Monkhorst-Pack k-mesh of 10 $\times$ 10 $\times$ 10 is used in all the AIMD simulations. A Nos{\'e} thermostat is applied at different temperatures (100 to 1500 K) for the various simulated cases, as required by the exchange-correlation functional to initiate atomistic displacements in the MoTaW CSA.

 A body-centered-cubic (BCC) supercell containing 18 atoms each of Mo, Ta and W is shown in Figure~\ref{fig:struct}(a). We perform AIMD simulations at seven different temperatures (100, 300, 500, 700, 900, 1100 and 1500 K), and monitor the displacements (\textit{x}, \textit{y}, \textit{z}) and forces (\textit{F$_x$}, \textit{F$_y$}, \textit{F$_z$}) experienced by each atom. The dataset consists of 70,000 different snapshots containing relevant displacements, forces, and energies for all atoms in the CSA. We employ a structural descriptor \textit{V}, as expressed in Eq.~\ref{morse_eqn}, to capture the interatomic interactions and analyze the chemical disorder in the alloy. In Eq.~\ref{morse_eqn}, \textit{x}, \textit{y}, \textit{z} are the normalized Cartesian coordinates (normalized with respect to maximum magnitude), \textit{$Z_i$} refers to the atomic number and $\gamma$ is the variance (typically 2.0 for a normal Gaussian distribution).

\begin{equation}\label{morse_eqn}
V_{(x,y,z)} =\sum_{i=1}^{N} Z _{i} \exp ( -[(x-x_{i})^2+(y-y_{i})^2+\\(z-z_{i})^2]/(2\gamma^{2}))
\end{equation}

\begin{figure}[t]
\centering
\includegraphics[width = 1.0\textwidth]{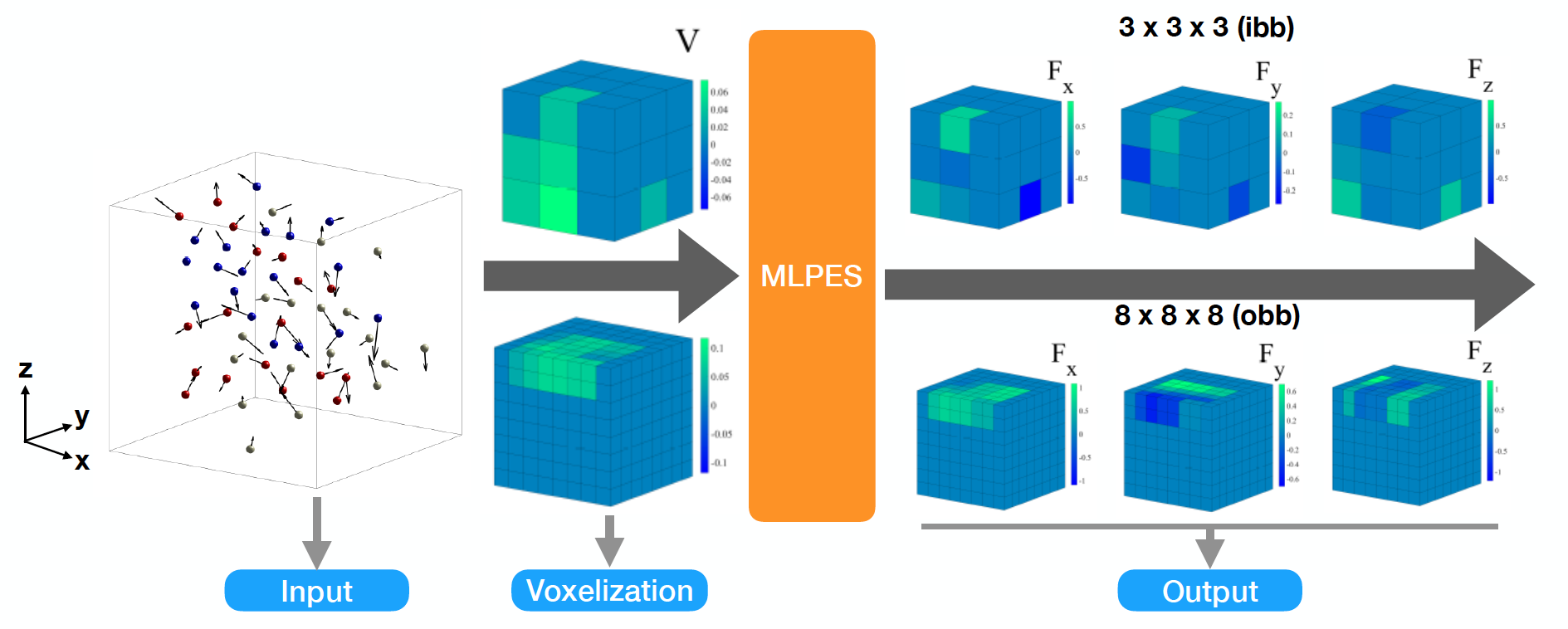}
\caption{Above representative atomic snapshot and corresponding voxel representation for the two approaches used in the voxel grid generation: 3 $\times$ 3 $\times$ 3 (ibb) and 8 $\times$ 8 $\times$ 8 (obb). The forces along the \textit{x}, \textit{y}, \textit{z} directions (\textit{F$_x$}, \textit{F$_y$}, \textit{F$_z$}) and the structure descriptor (V) can be visualized through the voxels.}
\label{fig:voxel}
\end{figure}

\subsection{Voxelization}
Typical AIMD output is extensive and complex for direct analysis with a deep neural network (DNNs). It is primarily very difficult because the data structure consists of a huge set of atomic coordinates and properties for all the atoms in the HEA. Hence, we use \textit{voxel} based representation of the atomic configuration, where the data is represented in a simple cellular structure bounded by a rectangular box. \textit{Voxel} is analogous to \textit{volume element}, and the generation of these elements is \textit{voxelization}, which generates a voxel grid with a user-specified grid resolution. Each voxel in the grid can contain none, one or more atoms depending on their \textit{x}, \textit{y}, \textit{z} coordinates. Voxel grid generation depends on an inside-outside test that compares the x, y, z coordinates of the atoms with respect to the grid boundaries of the voxels. The bounds are defined by the minimum and maximum points that can form the diagonal of the volume element, as illustrated in Fig. \ref{fig:struct}(b); the maximum and minimum points are represented by green spheres and other edge points are represented by orange spheres. We apply the following test to check whether a point of interest, \textit{P$_v$}, lies inside or outside the corresponding voxel. \textit{P$_v$} is considered inside the voxel if the inequalities in Eq.~\ref{Eqn:InsideOutside} are satisfied.
\begin{equation}
0 < \vec{k} \cdot \vec{u} \leq \vec{u} \cdot \vec{u}, \\
0 < \vec{k} \cdot \vec{v} \leq \vec{v} \cdot \vec{v}, \\
0 < \vec{k} \cdot \vec{w} \leq \vec{w} \cdot \vec{w}, \\
\label{Eqn:InsideOutside}
\end{equation}
where $\vec{u} = P_2 - P_1$, $\vec{v} = P_4 - P_1$, and $\vec{w} = P_5 - P_1$ are the basis vectors of the voxel illustrated in Figure~\ref{fig:voxel}  and $\vec{k} = P_v - P_1$. Once the atoms in simulation domain are decomposed to their respective voxels, the atomic properties are computed for each voxel. Here, each voxel contains the descriptors \textit{V}.

The voxel grid generation is implemented by two distinct approaches. In the first case (denoted as ibb for internal bounding box), we compute the bounding box of each configuration (a.k.a., snapshot or image) and subsequently generate the voxel representation based on that bounding box. In this way, each image is having with its own (internal) bounding box. We can say that the critical features of scale and relative interaction of atoms is clearly captured by this representation (because of its invariance with respect to the variance of scale in the atomic locations). We generate the voxel representation with resolution options of 4 $\times$ 4 $\times$ 4, 8 $\times$ 8 $\times$ 8, 16 $\times$ 16 $\times$ 16. The resolution of the voxel grid defines the number of volume elements bounded by the grid and, consequently, the size of the data to be processed.
In an alternate approach, we determine the bounding box of the entire set of 70,000 snapshots and generate voxel representation of each snapshot with respect to this outer bounding box (denoted as obb). As this representation is not invariant of the scale nor does it preserve the uniqueness of the atomic locations, it is important to have a very fine resolution of voxel grid. Physically, this could provide better generalization compared to ibb due to its accommodating the scale variance and with an additional cost of more computational resources. Physically, while the differences in atom locations are attributed to effect of kinetic energies in the AIMD simulations at the different temperatures, by using a proper voxel resolution all possible interactions within the alloy are included in the training dataset. Figure 2 illustrates a snapshot of the coordinates for the 54-atom ternary alloy structure, and their corresponding ibb and obb representations. 

We implement the entire voxelization process in Python programming language. To accelerate the voxelization operations, we employ Python's integrated multi-processing approach with \texttt{Numba} \cite{Lam2015}. A \texttt{first-in-first-out} (FIFO) \texttt{queue} is created using the queue package and all 70,000 configurations are added to the \texttt{queue}. We use the threading package using parallel programming to process the FIFO \texttt{queue} of the configurations. The voxelization subroutines are optimized and pre-compiled using the just-in-time compilation capability of \texttt{Numba} package. Applying optimization and pre-compilation steps before the subroutine execution allowed us to substantially (2 to 3$\times$) reduce the processing time spent for generating the voxel grids on a workstation equipped with Intel Xeon E5-2630 v3 processor and 32 gigabytes of RAM.
\begin{figure}[t]
\centering
\includegraphics[width = 1\textwidth]{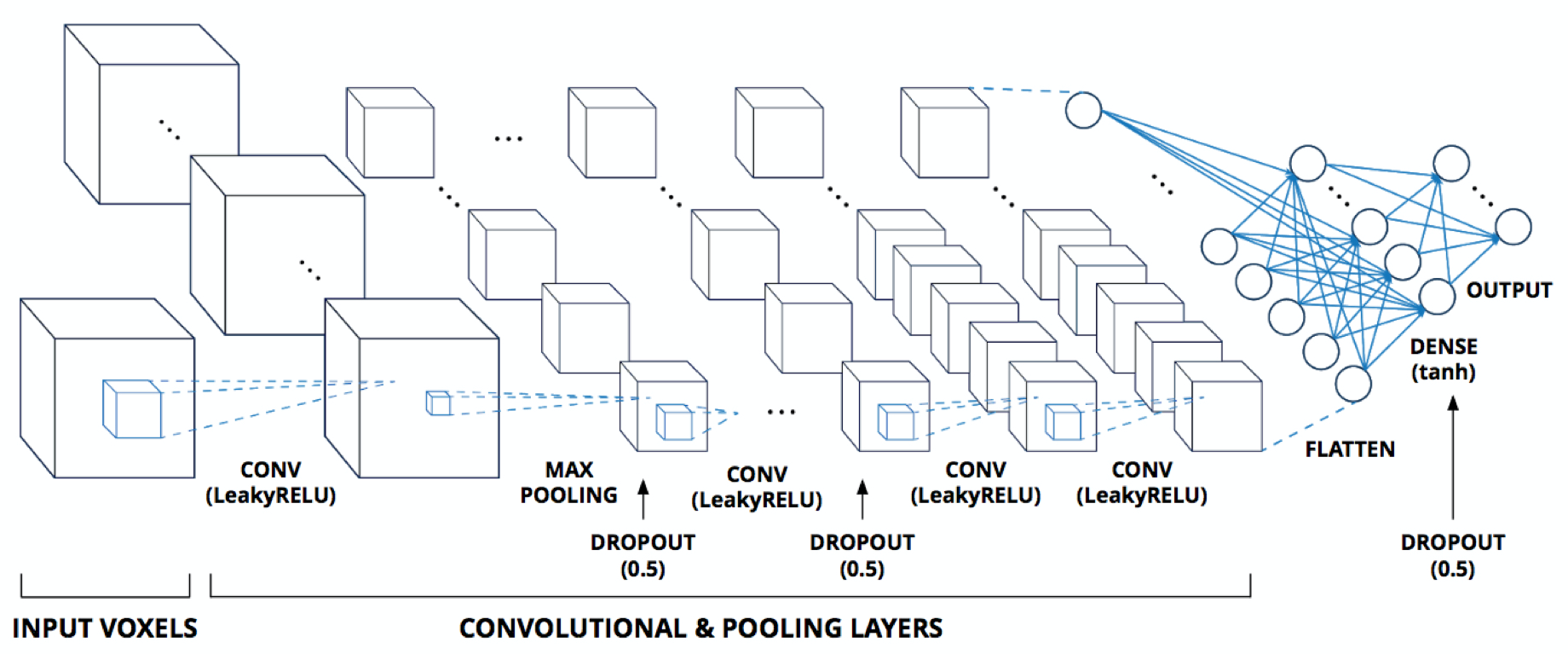}
\caption{A Convolutional Neural Network (CNN) architecture is used to machine learn the relationship between the PES and the atomic coordinates of the CSA. The CNN architecture primarily consists of convolutional layers followed by max-pooling layers and one fully-connected layer at the end succeeded by the output.}
\label{fig:CNN}
\end{figure}

\section{Deep Convolutional Neural Networks}\label{sec:DNN}
Data encountered in real life often involves learning from three dimensional (3D) data. Naturally, this has been a major area of research in itself since the eruption of Deep Learning. Many researchers have worked on using 3D convolutional networks (3DCNN) to learn from 3D data. The very first work was for the task of object detection from a 3D computer-aided design (CAD) geometry of simple objects (such as bed, or chair) using a 3DCNN (called as VoxNet)~\cite{wu20153d,ioannidou2017deep}. Further more, researchers have exploited this idea in several areas such as prediction from the point cloud data obtained from LIDARs~\cite{maturana}, engineering data used in design for manufacturing applications~\cite{sambitgmp}, and rendering smooth 3D graphics~\cite{tatarchenko2017octree}. 

A deep neural network is made up of several layers of connection forming one network which takes a input $x$ and produces an output $y$. Each connecting layer($l_i$) in the network can be represented as $y_{l_i} = \sigma(W_{l_i}.x_{l_i} + b_{l_i})$, where $\sigma(.)$ represents a non-linear activation function, $W_{l_i}$ and $b_{l_i}$ are the weights and biases, respectively, for connecting the input neurons($x_{l_i}$) to the output($y_{l_i}$) neurons. The connections could be as simple as a dense connection between every input neuron and output neuron in the layer. However, all connections in a dense connection layer may not be meaningful and the sample complexity to learn the connections would be high. A convolution connection instead of a dense connection helps in alleviating this issue.
The convolution operation ($\otimes$) is given by
\begin{equation}
W[m,n,p] \otimes x[m,n,p] = \sum_{i=-h}^{i=h}\sum_{j=-l}^{j=l}\sum_{k=-q}^{k=q} W[i,j,k].x[m-i,n-j,p-k]
\end{equation}

A series of convolutional connections, non-linear activations, pooling forms a convolutional neural network (CNN). The voxelized data obtained from the process explained in the previous section is fed into the CNN model. The network architecture, shown in Figure 3, comprises of multiple layers with max pooling and dropout in-between the layers. To account for the difference in the resolution of the ibb and obb, during the training of ibb and obb method, we modify the above parameters by reducing the filter size and no. of filters etc. For the obb methodology, the deep neural network consists of 3 layers with 256 filters and the kernel size of 2 $\times$ 2 $\times$ 2 in the first convolution layer. The strides are 1 $\times$ 1 $\times$ 1 in each convolution layer with a dropout of 0.5 in between the different layers. Subsequent layers had a 50\% reduction in filters, while kernel size and strides were constant. For the ibb methodology, increasing the filters and their size along with strides had negligible effect on the performance of the model. All the convolution layers have the LeakyReLu activation ($\alpha$ = 0.001) (except the final 3 dense layers, which had a tanh activation). Finally, the entire network is reduced to three fully connected layers, the layers bearing correspondence to \textit{F$_x$}, \textit{F$_y$} and \textit{F$_z$} obtained from the AIMD simulations.

\begin{figure}[t]
\centering
\includegraphics[width = 0.7\textwidth]{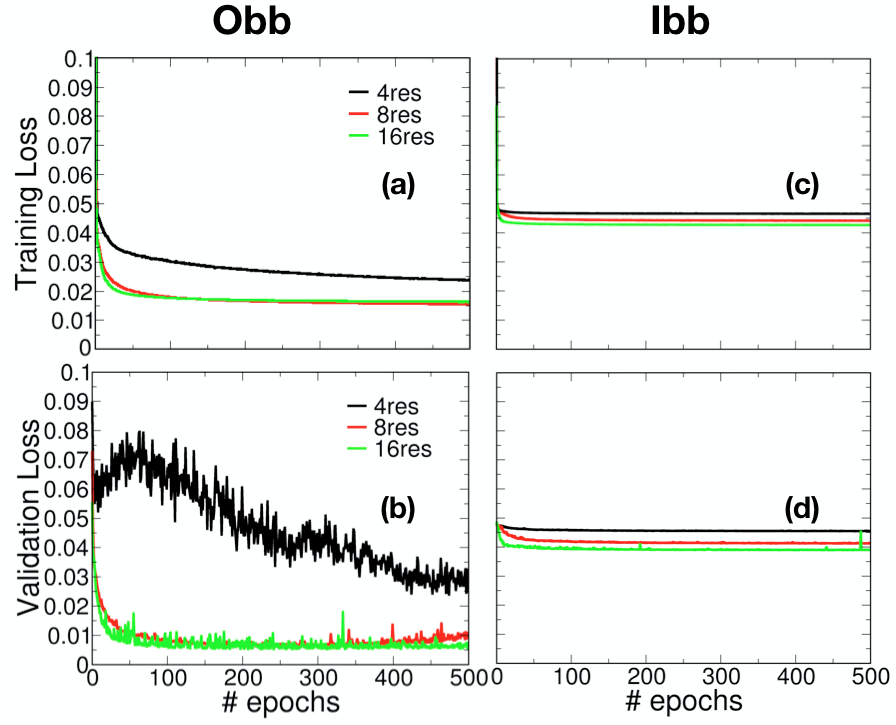}
\caption{Training (a,c) loss, and Validation (b,d) loss  is shown as function of epochs for both the outer bounding box (obb) and the inner bounding box (ibb) methodologies. The losses (Training and Validation) are presented for the different resolutions of the voxel grid representations (ibb and obb:  4 $\times$ 4 $\times$ 4, 8 $\times$ 8 $\times$ 8 and 16 $\times$ 16 $\times$ 16) of the 54-atom cubic simulation cell. The results indicate that the obb method is more effective than the ibb method. With increased resolution around 8 $\times$ 8 $\times$ 8, we have training and validation loss around 0.02 and 0.01, respectively. Increasing resolution beyond 8 marginally improves the performance (validation) for the obb methodology. For the ibb method, we find that increasing resolution improves performance but even for 16 $\times$ 16 $\times$ 16 resolution we have losses between 0.04 and 0.045 for both training and validation.}
\label{fig:results}
\end{figure}
\begin{figure}[h]
\centering
\includegraphics[width = 0.5\textwidth]{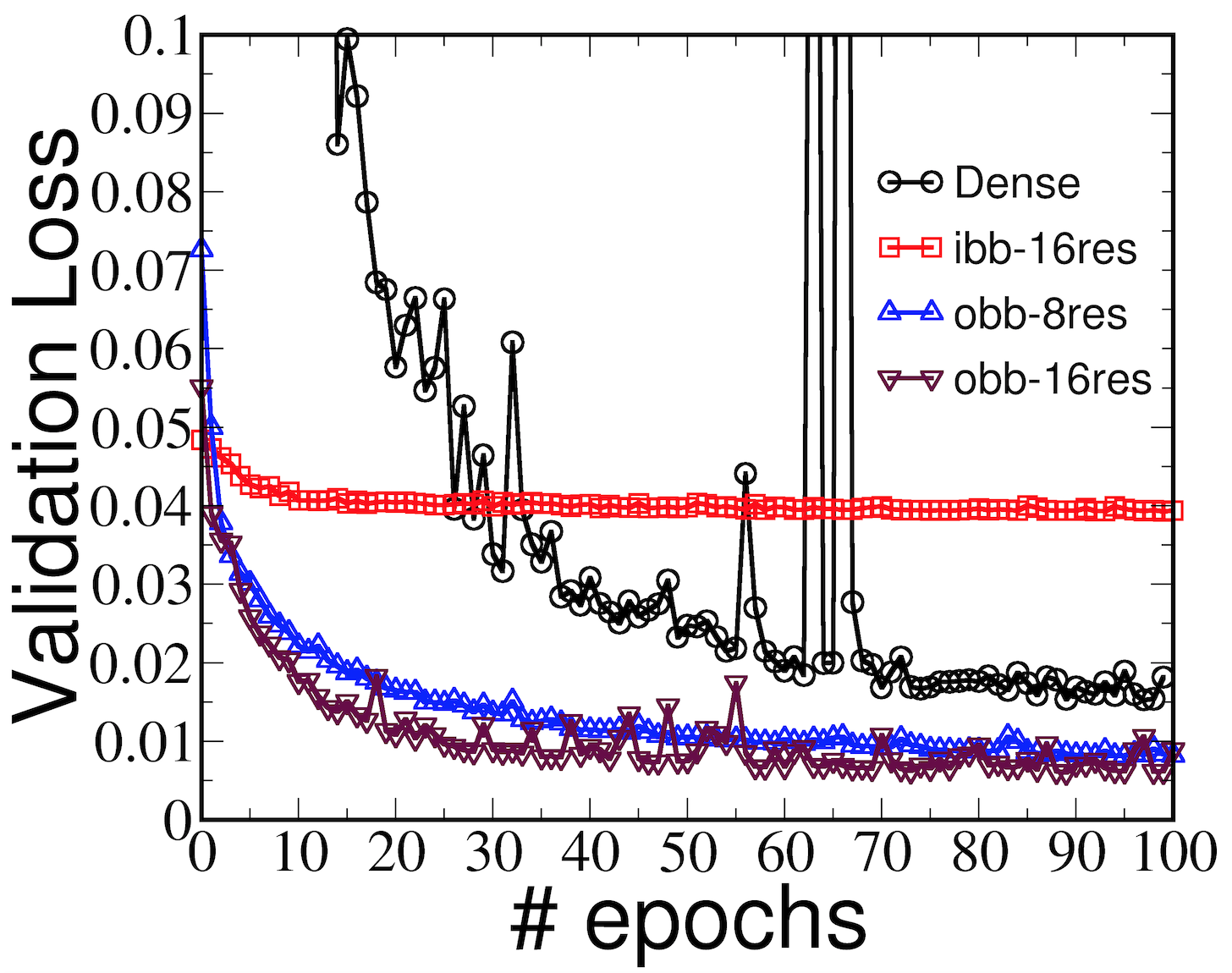}
\caption{A comparison of validation loss is presented for a dense network against the proposed CNN model employing the voxelized data (ibb: 16 $\times$ 16 $\times$ 16, and obb: 8 $\times$ 8 $\times$ 8 and 16 $\times$ 16 $\times$ 16) as the input. The results indicate that there is improvement in performance when using the obb implementation within the CNN model over traditional networks.}
\label{fig:results}
\end{figure}
\section{Results and Discussion}\label{sec:results}
For the present implementation, we use the Adam \cite{kingma2014adam} optimizer along with the L2-norm as the loss function. The test data was 0.3$\times$ the total data-set (70,000) used for validation, while the training/validation was done for 500 epochs. The analysis were carried out in a computing architecture with two Nvidia K20 Kepler GPUs.

In Figure 4, the training (Fig. 4(a) and (c)) and the validation (Fig. 4(b) and (d)) losses for the different resolutions of the voxel grids and the two implementations: ibb and obb, are reproduced. We observe that with an increase in the resolution, training and validation losses are reduced significantly for obb methodology till 8 resolution voxel grid. Thus, to minimize losses, the critical size of the voxel grid for the obb method must be greater than or equal to 8. Beyond 8 resolution, there is marginal improvement in accuracy on the expense of significant increase in computational time. Hence, for optimum performance the 8 resolution voxel grid needs to be selected in the obb methodology. From Figures 4(c) and (d), we note the effect of voxel grid resolution in the ibb methodology is less drastic than the obb method. Increasing resolution size improves accuracy and reduces losses but the overall performance is still poor in comparison to the obb method. The losses are constrained around 0.045 (training) and 0.04 (validation) even for the 16 resolution voxel grid size. In contrast, in the obb method we find the losses to be around 0.02 (training) and 0.01 (validation) for the 8 resolution voxel grid. 
\begin{figure}[ht]
\centering
\includegraphics[width = 1\textwidth]{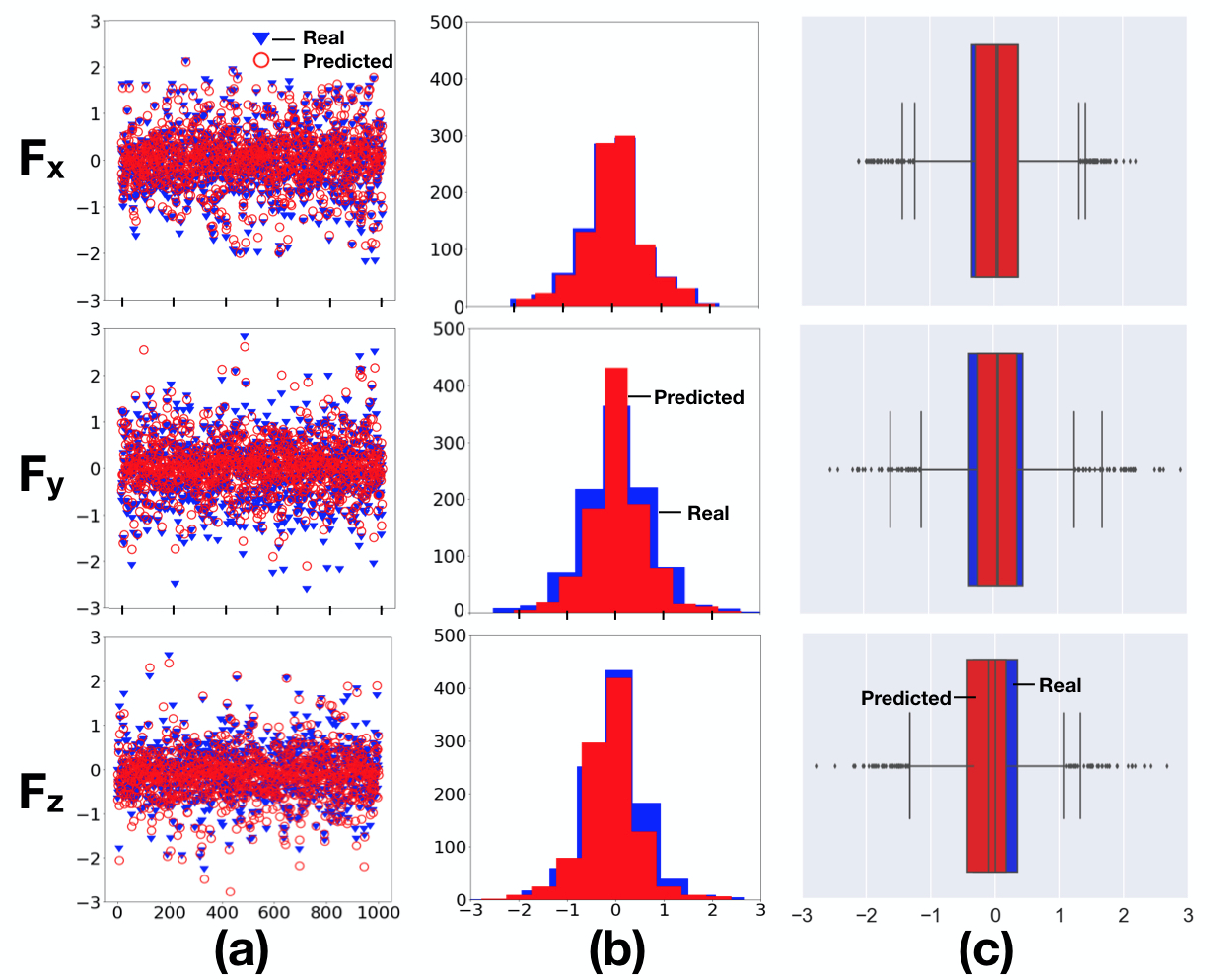}
\caption{For the 8 $\times$ 8 $\times$ 8 voxel resolution we evaluate the forces from our model against actual values. (a) shows the spread of the data (F$_{x}$,F$_{y}$, and F$_{z}$) while (b) indicates the distribution of the data (real and predicted) and (c) the median of the data alongside its 25\% and 75\% quantiles spread.   }
\label{fig:results}
\end{figure}
To show the effectiveness of the present implementation of using voxelization alongside convolutional neural network, we plot a comparison of validation loss (Fig. 5). It is observed that a fully connected deep network representing a dense configuration is out-performed by the CNN network (with obb voxelization framework). For the inner-bounding-box (ibb) voxelization methodology, the performance was poor in comparison to a full dense configuration, while for the outer-bounding-box (obb) methodology there occurs significant reduction in the losses, implying an enhanced accuracy for the proposed mocel. The losses (validation) for the dense network is around 0.02 while the obb 16 resolution model yields a loss around 0.0075. Thus, the inherent nature of the CNN to learn the local features more effectively and perform well even with a higher number of learnable parameters can be utilized to develop network architectures to generate the potential energy surface for complex multi-component alloy systems more effectively than the use of traditional dense networks.

The accuracy of the model is shown through the comparison between the actual values of forces and the predicted forces for a test set of coordinates. The comparison results are shown in Fig. 6 with blue marker representing the real force and red marker representing the predicted force. Figure 6(a) and 6(b) shows the distribution of the data points. In Fig. 6(a) the real forces in the three directions (x, y, z) are marked for 1000 test data points. The real forces distribution is largely concentrated around the center with few large values. Similar distribution is observed in all the three directions for the real force. We can observe in Figure 6(a) that the corresponding red markers (predicted values) are very close to the real forces, confirmed in Figure 6(b) (histogram), and Figure 6(c) (as Box plot). The distribution that resembles a Gaussian distribution is very similar for the real and predicted forces except for a few columns of data. Out of the three directional components of the force, F$_{x}$ (Force in x direction) is predicted most accurately. The box plot shown in Figure 6(c) correlates with this observation. The median and the (25, 75)$\%$ quantiles match exactly for F$_{x}$ while there are deviations in the values of F$_{y}$ and F$_{z}$ from the real forces. In case of F$_{y}$, the median of real data and predicted data are significantly close, however width of the box does not match for the two data sets. It shows that the variance in the real data set of F$_{y}$ is not captured perfectly by the model. Similarly we observe some discrepancies between the real and predicted values of F$_{z}$. In spite of these slight differences, this model provides a very good platform to create alternative solutions to the computationally expansive DFT-based simulations and model interactions between atoms.      

\section{Conclusions}
Using CNN to estimate the interatomic potential of CSAs is novel and previously only few attempts have been made to generate the potential for one elemental systems \cite{Artrith2016,Liu2017,Behler2016}. Prior studies have often discussed low accuracy and high validation errors associated with dense networks. Our results corroborate that using a CNN with voxelization can suitably capture the local environments as well as atomic specific properties like bond length, which eventually could be utilized to generate the inter-atomic potential or alloy specific properties. We find that increasing the voxelization resolution in the obb method offers significant reduction in losses (training and validation). With the ibb method, we find that the performance even with increasing voxel resolution is limited, as it fails to capture the local chemical environments and generalizes the results. Future efforts involves  directly comparing the structural and transport property predictions using ML potential with that from first-principles simulations.

\section*{Acknowledgements} 
Fruitful discussions on ML potential with  Dr. Nikolai Zarkevich at Ames Laboratory is greatly appreciated. The work at Ames Laboratory was funded by the U.S. Department of Energy (DOE),  Office of Science, Basic Energy Sciences, Materials Science and Engineering Division, and, in part (for A.S.) by Director's discretionary funds. Ames Laboratory is operated for the U.S. DOE by Iowa State University under Contract No. DE-AC02-07CH11358.  
The work was supported, in part, by the Office of Naval Research (ONR) through awards N00014-16-1-2548, N00014-18-1-2484 and by the U.S. AFOSR under the YIP grant FA9550-17-1-0220. Any opinions, findings and conclusions or recommendations expressed in this publication are those of the authors and do not necessarily reflect the views of the sponsoring agencies.
 \newpage
\bibliographystyle{unsrt}
\bibliography{references}
\end{document}